\documentclass[a4paper,11pt]{article}
\pdfoutput=1 

\usepackage{float}

\usepackage{jcappub} 

\usepackage[T1]{fontenc} 

\title{\boldmath Large scale structures and the cubic galileon model}

\author[a,b]{Sourav Bhattacharya,}
\author[a,c,d]{Konstantinos F. Dialektopoulos,}
\author[a]{Theodore N. Tomaras}

\affiliation[a]{ITCP and Department of Physics, University of Crete, 70013 Heraklion, Greece}
\affiliation[b]{Inter-University Centre for Astronomy and Astrophysics (IUCAA), Pune-411007, India\footnote[1]{Current affiliation}}
\affiliation[c]{Dipartimento di Fisica, Universit\'a di Napoli {}``Federico II'', Compl. Univ. di Monte S. Angelo, Edificio G, Via Cinthia, I-80126, Napoli, Italy\footnotemark }
\affiliation[d]{INFN Sezione  di Napoli, Compl. Univ. di Monte S. Angelo, Edificio G, Via Cinthia, I-80126, Napoli, Italy.}

\emailAdd{sbhatta@iucaa.in}
\emailAdd{dialektopoulos@na.infn.it}
\emailAdd{tomaras@physics.uoc.gr}

\abstract{\noindent
The maximum size of a bound cosmic structure is computed perturbatively as a function of its mass in the framework of the cubic galileon, proposed recently to model the dark energy of our Universe. Comparison of our results with observations constrains the matter-galileon coupling of the model to $0.033\lesssim \alpha \lesssim 0.17$, thus improving previous bounds based solely on solar system physics.}

\begin{document}
\maketitle
\flushbottom

\section{Introduction}
\label{sec:intro}

It is well known that so far the $\Lambda$-Cold Dark Matter ($\Lambda$CDM) model has passed in flying colours the cosmological observation tests, starting from the redshift of type Ia supernovae, the Hubble rate, the galaxy clustering and so on. The corresponding value of the cosmological constant, with a corresponding energy density determined to be of the order of $(10^{-3} eV)^4$, tiny compared to any other scale of particle physics - with the exception of the neutrino mass spectrum - has no satisfactory explanation to date. Interpreted as the vacuum energy of the fundamental theory of nature, it is just fine tuned to the above value since no symmetry has so far been discovered, which would result in a naturally small value for it. One might hope that supersymmetry could be useful in relation to the cosmological constant problem ~\cite{Martin:2012bt}, but still there is no experimental or observational evidence for it. In addition, no convincing dynamical mechanism has so far been presented, which would explain the present small value of $\Lambda$ as the evolution of an initially large value responsible for the inflationary period in the early universe, see e.g.~\cite{polyakov, ait, woodard}. As a result, we are instead facing the additional {\it coincidence puzzle} of why its corresponding energy density happened to be {\it today} of the same order of magnitude as the matter density in the Universe.

The above theoretical difficulty to attribute the observed anti-gravitational effects in nature to the workings of the rather mysterious cosmological constant, has triggered vigorous research for new gravity theories, which could describe satisfactorily the accelerating expansion of the Universe without the introduction of a cosmological constant, but, instead, by deviating from Einstein's General Relativity~\cite{review}. Such theories could predict time dependent dark energy density as opposed to $\Lambda$.
We refer our reader to~\cite{Sahni:2014ooa} for an interesting argument in favour of such dark energy candidates in a quite model independent way, based upon the Baryon Acoustic Oscillations data.    

Among such theories, the class of ``galileon models" were proposed as an alternative to the dark energy and have gained attention in recent years. Even though there are various kinds of fields which can in principle play the role of the galileon, the best understood paradigm and the one we shall focus on in what follows, is a scalar field $\phi$ coupled to gravity and with its action invariant under the shift symmetry in the flat spacetime, $\partial_a \phi\to \partial_a \phi + v_a $, where $v_a$ is a constant 1-form~\cite{Vikman, deffayet}. There can be up to fourth order derivative terms admitting this symmetry in flat spacetimes.  We refer our reader to~\cite{Martin-Moruno:2015kaa} and references therein for a discussion of how the galileon can serve as an attractor towards the de Sitter spacetime.

The galileon is a generalization~\cite{Nicolis1} of the so called Dvali-Gabadadze-Porrati massive gravity model~\cite{Dvali}-\cite{Rham}. As mentioned above, all such models contain non-linear derivative terms, which can describe the accelerated expansion of our universe at large scales. On the other hand, at small scales like e.g. the solar system, we must recover General Relativity in order to be consistent with the classic tests of Einstein's gravity. In order to achieve this, all such viable theories make use of the so-called ``Vainshtein mechanism" (see~\cite{babichev} for a review and also references therein), whose function is precisely to screen the effects of the higher derivative terms at small length scales. The length scale inside which this mechanism is effective is known as the ``Vainshtein radius". This feature of the model will play a crucial role in our study of the sizes of the galaxy clusters observed in our Universe. 

Since the galileon does not exclude a cosmological term from the gravitational action, it is fair to say that at the quantum level the galileon models have not resolved the fine tuning problem of the cosmological constant. Thus, they suffer from it exactly like the $\Lambda$CDM paradigm. Nevertheless, these models have attracted considerable interest recently and are being studied on their own right as alternative phenomenologically viable theories of gravity. Thus, in~\cite{Barreira:2013eea} one may find a nonlinear analysis on structure formation in the context of the cubic galileon model without any explicit matter-galileon coupling. Furthermore, aspects of weak gravitational lensing in galileon models can be found in~\cite{Tessore:2015sma}, while e.g.~\cite{Ali:2012cv, Hossain:2012qm, Bhattacharya:2015wlz} contain results on galileon physics with additional potential terms. Also, in the sector of stationary black holes, discussions on no hair theorem and their possible violations in the presence of the galileon can be found in e.g.~\cite{Hui:2012qt, Chris}.

The scalar galileon field couples to ordinary matter in order, in particular, to invoke the Vainshtein mechanism in the solar system. This results in non-geodesic motion for test particles. It is therefore interesting to study the phenomenology of the galileon-matter coupling parameter $\alpha$, which violates the weak equivalence principle. The effect of such coupling in the context of Mercury's perihelion precession have for cubic galleons been analysed in~\cite{Iorio:2012pv} and for quartic in~\cite{Andrews:2013qva}. In particular, the analysis of Ref.~\cite{Iorio:2012pv} constrains $\alpha$ to $|\alpha|\lesssim 0.3$. Further generalization of such galileon models with two scalar fields and their phenomenological implications can be found in~\cite{Padilla:2010de, Padilla:2010tj}.

In this paper we shall investigate the prediction of the cubic galileon~\cite{Vikman2} (see also~\cite{Babichev:2012re} and references therein) for the maximum sizes of large scale cosmic structures as a function of their mass and the matter-galileon coupling parameter $\alpha$. The maximum size will be determined by the maximum turn around radius, i.e. the point where the gravitational attractive force due to matter distribution and the repulsive force due to the dark energy balance each other acting on a test particle. For test particles without any angular momentum, this radius naturally gives an estimate of the maximum size a cosmic bound structure of a given mass can have. In other words, the maximum turn around length scale is perhaps the unique arena in gravitational physics, where both the attraction due to matter and the repulsion due to the dark energy are equally important. The relation of the maximum size determined by the turnaround radius (as opposed to the more commonly used virial radius) to the mass of a cosmic structure is a recently proposed quite reliable observable~\cite{Pavlidou:2013zha}, which can effectively distinguish the various alternative cosmological models~\cite{Tanoglidis:2014lea} and set constraints on their parameters~\cite{Pavlidou:2014aia, Bhattacharya:2015iha, Tanoglidis2015}. We also refer our reader to~\cite{Faraoni:2015zqa}, for a discussion on general calculation of the turn around radius in alternative gravity theories admitting McVittie solutions (e.g.~\cite{Kaloper:2010ec} and references therein).

The plan of this paper is as follows: In Section 2 we shall present the cubic galileon model, its field equations and the de Sitter attractor ``vacuum" solution, which will constitute the background of interest to us. In Section 3 we shall perturb the background by the introduction of a spherical overdensity and study the maximum turnaround radius as the distance from its center where the radial acceleration vanishes. We shall chiefly use the framework of the cosmological perturbation theory developed in~\cite{Pavlidou:2013zha, Pavlidou:2014aia, Tanoglidis:2014lea}, in the context of Einstein's gravity. Comparison with the observed values of size and mass of known bound structures, will lead to new bounds on the matter-galileon coupling $\alpha$ in Sec.~3, thus improving the existing ones based solely on solar system physics. We shall end with a brief summary.   

We shall use mostly positive signature for the metric ($-,+,+,+$) and will set $c=1$ throughout.

\section{The model}
\noindent
The action of the cubic galileon model we will be studying reads~\cite{Vikman2, Babichev:2012re}, 
\begin{eqnarray}
S=\frac{1}{4\pi G_N}\int d^4 x \sqrt{-g} \left[ \frac{R}{4}-\frac{k_2}{2}\left(\nabla_a\phi\right) \left(\nabla^a\phi\right)-\frac{k_3}{2M^2} (\nabla^a\nabla_a \phi)\, \left(\nabla \phi\right)^2 \right] + S_{\rm matter}\, (\Psi_m,\, \widetilde{g}_{ab} ) \nonumber
\\
\label{g1}
\end{eqnarray}
where $g_{ab}$ is the spacetime metric, $\phi$ is the galileon field, while $k_2$, $k_3$ and $M$ are parameters. The parameter $M$, in particular, is related to the characteristic  length
scale of the theory and indeed we shall see that the Hubble horizon is of the order of $\sim M^{-1}$ here.  
The generic matter field $\Psi_m$ couples minimally to the effective or physical metric $\widetilde{g}_{ab}=e^{2\alpha\phi}g_{ab}$, with $\alpha$ the matter-galileon coupling parameter. Due to this coupling, a test particle will fail to follow a geodesic with respect to the spacetime metric $g_{ab}$. Such coupling could be thought of as motivated from the Brans-Dicke theory written in the Einstein frame (e.g.~\cite{review}), where the conformal transformation in the Jordan frame 
generates such kind of non-minimal coupling between the metric and the scalar field, resulting in deviation from the geodesic motion. In particular, if we set $k_3=0$ in~(\ref{g1}), we recover the Brans-Dicke theory in the Einstein frame.

It is evident that we may absorb the parameters $k_2$ and $k_3$ in the redefinitions of $\phi$, $M$ and $\alpha$. However, we shall retain them for the time being for reasons that will soon become clear. 
The motivation behind our choice of the cubic galileon is the fact that it gives us a positive 
cosmological constant $\Lambda$, quite naturally, without needing to invoke it by hand, see~\cite{Babichev:2012re} and references therein for details.

\subsection{The equations of motion}
To derive the equation for $\phi$, note that the expressions for the matter energy-momentum tensor corresponding to the two metrics, $\widetilde{g}_{ab}$ and $g_{ab}$ are given, respectively, by $\widetilde{T}^{\rm matt.}_{ab}=-(2/\sqrt{-\widetilde{g}}) \left(\delta S_{\rm matter}/\delta \widetilde{g}^{ab}\right)$ and $T^{\rm matt.}_{ab}=-(2/\sqrt{-g}) (\delta S_{\rm matter}/\delta g^{ab})$. Using the chain rule for the functional derivatives, we find $T^{\rm matt.}_{ab}=e^{2\alpha \phi} \widetilde{T}^{\rm matt.}_{ab}$, from which we obtain $T^{\rm matt.}=e^{4\alpha \phi} \widetilde{T}^{\rm matt.}$. Using these, we conclude that
\begin{eqnarray}
\frac{\delta S_{\rm matter}}{\delta \phi}=\alpha \sqrt{-\widetilde{g}}\, \widetilde{T}^{\rm matt.}=\alpha \sqrt{-g}\, T^{\rm matt.}.
\label{g2}
\end{eqnarray}
Then we get the following equation of motion for $\phi$
\begin{eqnarray}
\nabla_a\left[k_2 \nabla^a \phi-\frac{k_3}{2M^2} \nabla^a(\nabla \phi)^2 +\frac{k_3}{M^2} (\nabla^b\nabla_b\phi) \nabla^a \phi   \right]=-4\pi \alpha G_N T^{\rm matt.}.
\label{g3}
\end{eqnarray}
The vector field 
$$
J^a=k_2 \nabla^a \phi-\frac{k_3}{2M^2} \nabla^a(\nabla \phi)^2 +\frac{k_3}{M^2} (\nabla^b\nabla_b\phi) \nabla^a \phi
$$ 
appearing in~(\ref{g3}) is called the galileon 4-current, which in the absence of matter or for conformally invariant fields is covariantly conserved, $\nabla_aJ^a=0$.

Similarly, making use of the relations above, we obtain the equation of motion for the metric $g_{ab}$,
\begin{eqnarray}
G_{ab} =8\pi G_N \left[T_{ab}(\phi) +T^{\rm matt.}_{ab}\right],
\label{ga1}
\end{eqnarray}
where $T_{ab}(\phi)$ is the galileon energy-momentum tensor given by,
\begin{eqnarray}
8\pi G_N T_{ab}(\phi)= k_2[2(\nabla_a\phi)(\nabla_b\phi)-g_{ab}(\nabla \phi)^2]+\frac{k_3}{M^2}[2(\nabla_a\phi)(\nabla_b\phi) \nabla_c\nabla^c \phi + g_{ab} (\nabla_c\phi)\nabla^c(\nabla\phi)^2 \nonumber \\
- (\nabla_{(a}\phi )\nabla_{b)} (\nabla\phi)^2]. \nonumber \\
\label{g5}
\end{eqnarray}
%
First, we shall deal with the Friedmann-Robertson-Walker cosmological spacetime. Assuming, in accordance with observations, a spatially flat homogeneous background 
$$ds^2=-dt^2+a^2(t)\left[dx^2+dy^2+dz^2\right]~~{\rm and}~~\phi\equiv \phi(t),$$ 
the field equations~(\ref{g3}), (\ref{ga1}) read
\begin{eqnarray}
\partial_t\left(a^3(t)J_t \right)&=& -4\pi \alpha G_N a^3(t)  (\rho(t)-3P(t)),\nonumber\\
3{\cal H}^2 &=& 8 \pi G_N \rho(t) + k_2 \dot{\phi}^2 -\frac{6 k_3 \dot{\phi}^3 {\cal H}}{M^2},\nonumber\\
\frac{3\ddot{a}}{a}&=&-4\pi G_N \left(\rho(t)+3P (t)\right) +\dot{\phi}^2\left(2k_2 -\frac{3 k_3 \dot{\phi} {\cal H}}{M^2}  +\frac{3k_3 \ddot{\phi} }{M^2}\right),
\label{gad1}
\end{eqnarray}
where $J_t=k_2\dot{\phi}-3{\cal H} k_3 \dot{\phi} ^2/M^2$,
${\cal H}=\dot {a}/a$ is the Hubble rate, $\rho(t)$ and $P(t)$ are the density and pressure corresponding to any other matter fields, $T_{ab}^{\rm matt.}$ in Eq.~(\ref{ga1}). All spatial components of the galileon current vanish identically due to the symmetry.

\subsection{The de Sitter solution}
Since the objective of the galileon is to mimic the dark energy, we should first check whether primarily it can give 
an expanding solution at all i.e. ${\cal H}>0$, even in the absence of any other matter field. For $\rho(t)=0$, the second of Eqs.~(\ref{gad1}) gives
%
%
%
\begin{eqnarray}
{\cal H}= -\frac{k_3 \dot{\phi}^3}{M^2}\pm  \frac{k_3 \dot{\phi}^3}{M^2} \left[1+\frac{k_2 M^4}{3k_3^2 \dot {\phi}^4}\right]^{\frac12}.
\label{gad6}
\end{eqnarray}
Thus we must have $k_2 M^4/3k_3^2 \dot {\phi}^4\geq -1$, in order for ${\cal H}$ to be real. Then the above solution can have expanding branches, for example, if we take
$ k_3 \dot{\phi} <0$, and the `-' sign. If we consider additional matter field, it will increase the Hubble rate as long as $\rho(t)$ is positive, the second of Eqs.~(\ref{gad1}). 

For example, for the `dust' we have $\rho(t)=\rho_0/a^3(t)$ (with $a(t)\sim t^{\frac23}$), where $\rho_0$ is a constant. For any other matter with an equation of state with parameter $w>0$, the density $\rho(t)$ will decay faster than this with respect to $a(t)$. The first of Eqs.~(\ref{gad1}) now gives
\begin{eqnarray}
k_2\dot{\phi}-\frac{3{\cal H} k_3 \dot{\phi} ^2}{M^2}=-\frac{4\pi \alpha G_N \rho_0 t}{a^3(t)}+ \frac{C_1}{a^3(t)},
\label{gad4}
\end{eqnarray}
where $C_1$ is an integration constant.
Since we have already argued that `turning on' the galileon should increase ${\cal H}(t)$, the scale factor $a(t)$ appearing above should be faster than $t^{\frac23}$. Consequently, both the terms on the right hand side of the above equations vanish
at late times, giving
\begin{eqnarray}
{\cal H} = \frac{k_2 M^2 }{3k_3 \dot{\phi} },
\label{gad5}
\end{eqnarray}
an alternative expression for  ${\cal H}$ to that of~(\ref{gad6}).
Thus we have an expanding solution whenever $k_2 k_3 \dot {\phi} > 0$. Moreover, when $\dot{\phi}$ is a constant, we clearly have a de Sitter space. In other words, for the linear ansatz~\cite{Babichev:2012re} 
\begin{equation}
\phi=\phi_0+\phi_1 t\,,
\label{phi-ansatz}
\end{equation}
where $\phi_0$ and $\phi_1$ are constants, we get the de Sitter space with ${\cal H}=k_2M^2/(3 k_3 \phi_1)$. 

The de Sitter solution thus found acts as an attractor, at least if we assume a perturbative expansion in the coupling parameter $\alpha$. This can be seen by perturbing $\phi(t) \to \phi(t)+\delta \phi(t)$ by invoking some matter density $\delta \rho(t)$ in Eq.~(\ref{g3}). From our previous arguments, the resulting perturbation in the current, $\delta J_t$, behaves as $\sim (\alpha t+C_1)/a^3(t)$. Now, up to ${\cal O}(\alpha)$, we may take the scale factor to be the background de Sitter, resulting in a very fast decay. It is also clear that the contribution to the galileon energy-momentum tensor due to $\delta\phi(t)$ thus found will be ${\cal O} (\alpha^2)$ times a function decreasing with time.  
 
We now substitute the linear solution~(\ref{phi-ansatz}) into~(\ref{g5}) to get
\begin{eqnarray}
8\pi G_N T_{ab}(\phi)= +k_2 \phi_1^2 g_{ab},
\label{g6}
\end{eqnarray}
so that in order to get the equation of state for the cosmological constant, we must have $k_2<0$, and the de Sitter 
Hubble rate is given by
\begin{eqnarray}
{\cal H}^2= \frac{|k_2|\phi_1^2}{3} \equiv \frac{\Lambda_G}{3}
\label{g6'}
\end{eqnarray}
with $\Lambda_G$ the cosmological constant of this theory. Putting the above equation together with Eqs.~(\ref{gad5}),~(\ref{phi-ansatz}), we find that ${\cal H}\sim M$, thereby $M$ determining the characteristic length scale of the theory, as we stated below Eq.~(\ref{g1}).

However, $k_2<0$ gives a wrong sign 
in the kinetic term of~(\ref{g1}), apparently indicating ghost-like instability at very large length scales. This length scale is determined by the Vainshtein radius~\cite{Vikman2}. However, as was shown in that reference (see also~\cite{Babichev:2012re}), the perturbation of both the metric (due to matter) and the galileon field mix the spin-0 and spin-$2$ states. A diagonalization procedure then yields an effective metric through which the pure scalar (spin-$0$) mode propagates. It is the correct Lorentzian signature of the effective metric that  ensures that the matter perturbation does not become superluminal and the stability in maintained. It turns out that the perturbation is perfectly stable inside the Vainshtein radius, whereas in order to investigate stability outside it, seemingly one needs to take into account the contribution of the homogeneous cosmological matter density into the Hubble rate, but it is not certain whether the theory remains stable via this prescription.

Nevertheless, we shall see below that as long as the matter-galileon coupling is `small' but non-vanishing,  the maximum turn around radius is always located inside the Vainshtein radius (see also~\cite{Silva} for a discussion on stability of structures in a Brans-Dicke extension of the galileon with a `wrong' sign of the kinetic term, but without any reference to the maximum sizes of the structures). So, before prolonging the discussions on the stability, let us explicitly calculate the turn around radius first.

\section{The maximum turn around radius}
\noindent
We shall now introduce spatial inhomogeneity into the de Sitter spacetime thus found, in order to describe and estimate the maximum sizes of the large scale cosmic structures. As we have discussed in Sec.~1, this will be determined by the maximum turn around radius. For that, we shall study the dynamics of a test fluid outside a cosmic structure formed by non-relativistic or cold dark matter~\cite{Pavlidou:2014aia}. We shall assume the structure spherical, in which case all its mass can be taken to be at the centre, $r=0$. The size of a structure will be taken to be much large compared to its Schwarzschild radius. 
 In addition, we shall ignore any background homogeneous matter density contribution outside the structure, which is the region of interest for our purpose. Finally, we shall restrict ourselves to linear order in the matter-galileon coupling $\alpha$, as in the previous section and ignore galileon's backreaction on the spacetime metric. Nevertheless, since such coupling breaks the weak equivalence principle, it would affect the motion of a test fluid and hence, as we shall see, the turn around radius.  

The situation is described by the McVittie spacetime (see e.g.~\cite{Kaloper:2010ec} and references therein, also~\cite{Kastor:1992nn}), for which the metric $g_{ab}$ reads, when we are much outside the Schwarzschild radius of the body,
\begin{eqnarray}
ds^2=-(1+2\Phi(r,t))dt^2 +a^2(t)(1-2\Phi(r,t))\left[dx^2+dy^2+dz^2\right],
\label{g8}
\end{eqnarray}
where $a(t)=e^{{\cal H} t}$, $r^2=x^2+y^2+z^2$ and $\Phi(r,t)$ is the spherical gravitational potential generated by the central mass. The components of the Einstein tensor for this metric are given by 
\begin{eqnarray}
G_{00}=3{\cal H}^2+2{\nabla}^2 \Phi-6{\cal H} \dot{\Phi},~
G_{ij}=\left[-\frac{2{ \ddot a}}{a}+ {\cal H}^2 +\frac{8 {\ddot a} \Phi}{a}+ 2 \ddot{\Phi} -4{\cal H}^2\Phi+6 {\cal H} \dot{\Phi} \right]\delta_{ij}.
\label{g9}
\end{eqnarray}
For the source of the inhomogeneity to be a point mass at rest at $r=0$, the relevant energy-momentum tensor is $T^{\rm matt.}_{tt}=m\delta^3 (\vec{r} a(t) e^{\alpha \phi})$, $T^{\rm matt.}_{t i}=0=T_{ij}^{\rm matt.}$. Note that in the argument of the delta function, we have kept the full physical proper length appropriate in the frame of $\widetilde{g}_{ab}=e^{2\alpha \phi} g_{ab}$, in accordance with~(\ref{g1}). Clearly, as far as the physical mass is concerned, we must integrate 
$T^{\rm matt.}_{tt}$ over the proper spatial volume element $\equiv e^{3\alpha \phi} a^3 (t)d^3 x$, giving $m$. However, we note also that in the frame of $g_{ab}$, integration of $T_{tt}$ yields a time dependent mass function, $m(t)\equiv m/e^{3\alpha \phi}$.  

The length scale of the structures are essentially much smaller than the Hubble radius and hence the perturbation we are studying is subhorizon. Accordingly, we can ignore the temporal derivatives of the potential $\Phi(r,t)$ compared to its spatial derivatives. Then, while the spatially homogeneous part of the Einstein-galileon equations with $k_2<0$ gives the de Sitter universe as described in the previous section, the inhomogeneous perturbation gives the Poisson equation,
\begin{eqnarray}
{\nabla}^2 \Phi=4\pi G_N m \delta ^3 (\vec{r} a(t)e^{\alpha \phi} ).
\label{g10}
\end{eqnarray}
%
Integration of the above equation yields Newton's potential $\Phi=-G_N m /(e^{3\alpha \phi} ra(t))\ll1$, with an effective time dependent mass function $m(t)\equiv m/e^{3\alpha \phi}$, in accordance with the preceding discussions. Clearly,
for $\alpha=0$ this represents nothing but the Schwarzschild-de Sitter spacetime far away from the Schwarzschild radius written in the McVittie frame. 

Certainly, the galileon will also receive  a correction of ${\cal O}(\alpha)$ due to the central mass, Eq.~(\ref{g3}). However, this will give an ${\cal O}(\alpha^2)$ term via the matter-galileon coupling, into the physical metric $\widetilde{g}_{ab}=e^{2\alpha \phi} g_{ab}$. Since we are restricting   
ourselves to linear order in $\alpha$ throughout, we shall ignore such correction.

 Let us now study the dynamics of a non-relativistic test fluid with density $\delta \rho(t,x^i)$. 
Let the spatial velocity of the fluid with respect to the frame of $g_{ab}$ be $v^i=dx^i/dt$. Since the fluid is assumed to be non-relativistic, this velocity will be `small'. We take the energy-momentum tensor of the total fluid to be, to ${\cal O} (v^i)$,
\begin{eqnarray}
T^{t}{}_{t}=-\rho-\delta \rho,~~T^{i}{}_{t}=-\rho v^i,
\label{g11}
\end{eqnarray}
where $\rho$ and $\delta \rho$ correspond respectively to the energy density of the central  effective  point mass and the test fluid. Eq.~(\ref{ga1}) shows that we can use the conservation equation for the matter field with respect to $g_{ab}$, yielding
\begin{eqnarray}
\delta\dot{\rho}+\rho \partial_iv^i+3{\cal H} \rho=0~~{\rm and}~~
\dot{v}^i+2{\cal H} v^i+\frac{\partial_i\Phi}{a^2}=0.
\label{g12}
\end{eqnarray}
The proper length which a test fluid element experiences in the physical metric $\widetilde{g}_{ab}$ is given by, $\widetilde{r}^i=a(t) e^{\alpha \phi}r^i$, with $r^i=x,y,z$. We compute the proper
or physical velocity and acceleration for this length by differentiating $ d\widetilde{r}^i$ with respect to $d	\tilde{t}=e^{\alpha \phi}dt$, 
\begin{eqnarray}
\frac{d \widetilde{r}^i}{d\tilde{t}}&=&\frac{1}{e^{\alpha \phi}}\frac{d \widetilde{r}^i}{ dt}= \left[\dot{a}(t) r^i+\alpha a(t)\phi_1 r^i+a(t)v^i\right]\\
\frac{d^2 \widetilde{r}^i}{d \tilde{t}^2}&=& e^{-\alpha \phi}a(t)\left[\frac{\ddot{a}}{a}r^i+(2{\cal H} v^i+\dot{v}^i)+\alpha {\cal H} \phi_1 r^i \right],
\label{g13}
\end{eqnarray}
where in the last equation we have ignored second order term containing both $\alpha$ and $v^i$.
 The turn around radius is defined 
as the point where the physical acceleration vanishes, $d^2 \widetilde{r}^i/d{	\tilde{t}}^2=0$. 
%
%
We use the second of Eqs.~(\ref{g12}) into~(\ref{g13}) and multiply it by $e^{\alpha \phi}$. Also using spherical symmetry $\widetilde{r}^i=\widetilde{r}$ and substituting $\Phi=-G_N m/(ra(t) e^{3\alpha \phi})$, we obtain
%
%
%
\begin{eqnarray}
\frac{\ddot{a}}{a} \widetilde{r}-\frac{G_N m}{\widetilde{r}^2}+\alpha {\cal H} \phi_1  \widetilde{r} =0,
\label{g15}
\end{eqnarray}
which, using $a(t)= e^{{\cal H}t}$ gives the maximum turn around radius to ${\cal {O}}(\alpha)$,
\begin{eqnarray}
\widetilde{R}_{\rm TA,max.}\approx\left(\frac{G_N m}{{\cal {H}}^2 }\right)^{\frac13}\left(1-\frac{\alpha \phi_1}{3\cal H} \right).
\label{g16}
\end{eqnarray}
If we set $\alpha=0$ above, we recover the maximum turn around radius for the $\Lambda CDM$ model~\cite{Pavlidou:2014aia}.
Using Eqs.~(\ref{gad5}), (\ref{phi-ansatz}) and (\ref{g6'}) the above equation becomes
\begin{eqnarray}
\widetilde{R}_{\rm TA,max.}=\left(\frac{3G_N m}{\Lambda_G}\right)^{\frac13}\left(1-\frac{\alpha}{\sqrt{3} |k_2|^{\frac12}}\right).
\label{g17}
\end{eqnarray}
This expression gives the effect of the matter galileon coupling upon the turn around radius. Now, the analysis with the Einstein gravity made in~\cite{Pavlidou:2014aia} shows that for the largest cosmic structures like galaxy clusters with masses as large as $10^{15}M_{\odot}$ (e.g., the Virgo cluster,  $M\sim 10^{15} M_{\odot}$), the maximum turn around radii are only about $10\%$ larger than their actual observed sizes. In other words, any theory predicting 
the maximum sizes of those particular cosmic structures can be at most $10\%$ less than the prediction of $\Lambda$CDM.
We now set $k_2=-1$. Clearly, this corresponds to the scaling $\phi\to |k_2| \phi$, $\alpha \to \alpha/|k_2|$ and $k_3\to k_3/|k_2|^{\frac32}$ in the action,~(\ref{g1}). If we take $\Lambda_G$
above to be the $\Lambda$ of Einstein's theory, assume $\alpha$ to be positive, we find from the above criterion of stability that $\alpha <0.17 $. We cannot conclude anything specific if $\alpha$ is negative, from this discussion.

But there is still a caveat in the stability issue -- i.e. the Vainshtein radius which we have not addressed yet explicitly. Our prediction of the maximum turn around radius can be safely trusted only if it lies inside that radius, as discussed at the end of the previous section. For this model the expression for the Vainshtein radius is given by, with $k_2=-1$~\cite{Babichev:2012re},
\begin{eqnarray}
R_V=\left[\frac{8k_3 G_N m}{ M^2}\left(\alpha + \frac{k_3 \phi_1^2}{M^2}\right)  \right]^{\frac13}.
\label{g18}
\end{eqnarray}
The above radius is written in the frame of $g_{ab}$. However, note that in order to get the effect of the matter-galileon coupling, we must take the mass $m$ appearing above to be time dependent,  $m(t)\equiv m \,e^{-3\alpha \phi}$, as pointed out earlier. Then we have in the physical frame
\begin{eqnarray}
\widetilde{R}_V=\left[\frac{8k_3 G_N m}{ M^2}\left(\alpha + \frac{k_3 \phi_1^2}{M^2}\right)  \right]^{\frac13}.
\label{g19}
\end{eqnarray}
In order to compare this with $\widetilde{R}_{\rm TA,max.}$, we use Eqs.~(\ref{gad5}), (\ref{g6'}) to get
\begin{eqnarray}
\widetilde{R}_V\approx\left(\frac{3 G_N m}{\Lambda_G}\right)^{\frac13}\left(1 + \frac{\alpha (1-1/27)}{\sqrt{3} }-\frac{1}{27} \right)=\widetilde{R}_{\rm TA,max.}\left(1 + \frac{53 \alpha}{27 \sqrt{3} }-\frac{1}{27} \right).
\label{g20}
\end{eqnarray}
Thus, the Vainshtein radius is larger than the maximum turn around one for $\alpha\gtrsim 0.033$. Note, in particular, that for $\alpha \leq 0$ the $\widetilde R_{\rm TA,max}$ lies in the potentially unstable regime, while for $\alpha=0.17$ $\widetilde R_{\rm V} \simeq 1.16 \widetilde R_{\rm TA,max}$. In other words, the consideration of the Vainshtein radius 
determines the sign of $\alpha$ and also offers us a lower bound. Note also that the upper bound we have found on $\alpha$ is about $50\%$ improvement than that found from the solar system phenomenology, $\alpha \lesssim 0.3$~\cite{Iorio:2012pv}.

We shall now make a plot of the various parameters of the theory based on the bound $0.033\lesssim \alpha \lesssim 0.17$.
Let $\Lambda_0$ ($\sim 10^{-52}m^{-2}$) be the value of the cosmological constant in $\Lambda$CDM.  We write $\Lambda_G= g^3 \Lambda_0$, $g$ being a dimensionless real number. 
Let $\widetilde{R}^0_{\rm TA,~max.}$ be the maximum turn around radius in $\Lambda$CDM and let us also define a dimensionless
number, $q={\widetilde{R}_{\rm TA,max.}/\widetilde{R}^0_{\rm TA,max.}}$. Putting these all in together, Eq.~(\ref{g17}) becomes 
\begin{eqnarray}
q= \frac{1-\frac{\alpha}{\sqrt{3}}}{g},
\label{g21}
\end{eqnarray}
where $\alpha=0$ and $g=1$ gives $q=1$ corresponding to $\Lambda$CDM.
We have a minimum value of $\alpha= 0.033$, which for $g=1$, gives $q=0.98$. Using these ranges, we determine how 
much $\Lambda_G$ could be bigger than $\Lambda_0$, in Fig.~1.
\begin{figure}[H]
\centering
\rotatebox{0}{
\includegraphics[height=4.5cm, width=6cm]{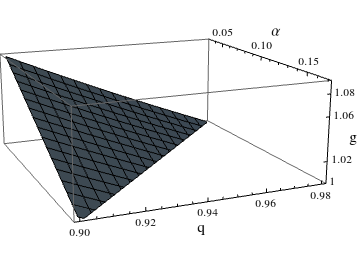}}
\caption{Variation of $g$ as of Eq.~(\ref{g21}) -- we find $ g \leq 1.09$.}
\label{fig:test}
\end{figure}
%

\section{Summary and outlook}
\noindent
We have studied the maximum turnaround radius in the context of the cubic galileon model in the presence of a weak non-vanishing matter-galileon coupling.
We used the de Sitter vacuum background space of this theory and the matter-galileon coupling, in order to estimate perturbatively the maximum allowed sizes of cosmic structures. The observational data and the requirement that the structures should lie inside the Vainshtein  radius gave us a bound on the matter-galileon coupling constant $\alpha$, $0.033\lesssim \alpha \lesssim 0.17$, thereby improving the existing bound $|\alpha| \lesssim 0.3$~\cite{Iorio:2012pv} found on the basis of solar system phenomenology. While it is not completely clear whether this theory is free from ghost instability outside the Vainshtein radius, our analysis guarantees the structures to remain inside it, and hence their stability. Hopefully this is a step towards understanding this particularly simple model of the galileon family.   

We note a possible weakness~\cite{Kaloper:2010ec} of the McVittie spacetime we used here. Precisely, all McVittie descriptions take the background matter density to be homogeneous, $\rho\equiv\rho(\tau)$, in addition to the usual, central overdensity. But it is natural to expect the overdensity to induce some inhomogeneity in the background density as well, due to the breaking of the maximal spatial symmetry. While this could be a possible concern in the strong gravity regime, where we are close to the Schwarzschild radius, in the framework of the cosmological perturbation theory such as our current analysis, where we are much outside the Schwarzschild radius of the central object, the McVittie description may well be a reasonable approximation.  

 Before we end, we mention here a plausible way to deal with the possible ghost instability of this theory outside the Vainshtein radius (Sec.~2.2), by putting in some extra ingredients.  Precisely, let us multiply the kinetic term in~(\ref{g1}) with some $\zeta(\phi)$ and add a potential $V(\phi)$ as well. These additions should be such that firstly, $\zeta(\phi)$ is $+1$ and $V(\phi)\to 0$ inside the Vainshtein radius, so that the theory is effectively the same as what we have considered so far. On the other hand at very large  length scales, we must have $\zeta(\phi)<0$ in order to avoid the ghost instability. Since the largest scales are probed only by `low' momenta, we expect only the quadratic term in~(\ref{g1}) to be functional there. In that case it is easy to see that the theory fails to satisfy the strong energy condition ($\equiv \rho+3P <0$, necessary to drive the accelerated expansion) unless we add an appropriate potential, too.  A reasonable question would then be, why we are interested in this particular theory, instead of some other dark energy model with a potential, such as the quintessence~\cite{Ratra}? This is because the cubic galileons, as we have seen, could mimic the $\Lambda{\rm CDM}$ model very well, but with a very important qualitative departure -- the deviation from the geodesic equation (Sec.~3). Such departures are always interesting in their own right, in order to distinguish how {\it fundamentally} different a model is, from the $\Lambda{\rm CDM}$. Nevertheless, this proposal needs to be actually demonstrated to work well, which we plan to do in a future publication.

Perhaps the necessity of the above extension is further emphasized when we consider~(\ref{g1}) from a standard field theoretic point of view,
as the theory is expected to become strongly coupled above its characteristic mass scale, $M$. For a large class of galileon models
with potentials, it has been shown recently in~\cite{Kunimitsu:2015faa}, using effective action calculations  for a slow role inflation, that this apparent feature is {\it not} indeed the case -- the actual strong coupling scale is much higher than the characteristic mass scale.  The particular theory we are addressing currently, needs to be checked in that way as well, after we find its appropriate extension with potential to avoid the ghost instability, as proposed in the preceding paragraph.

 An important question now would be : why were we so much interested in making the maximum allowable sizes of structures to be smaller than the Vainshtein radius, provided there is indeed some way to avoid the ghost instability outside it? Note that this means $\alpha<0$ in~(\ref{g20}), which would give no constraint on it, whatsoever. However, this would lead to a new problem as follows. On or in an infinitesimal neighborhood of the Vainshtein radius, the galileon field is expected to become non-perturbative~\cite{Babichev:2012re}.
Consequently, the equation of motion for a massive test particle, $u^a\nabla_au_b=\alpha \nabla_b \phi$, gets a non-perturbative contribution from the galileon field there, too. Had $R_{\rm TA,~Max}$
indeed been larger than $R_V$, the actual surface of the structures we have considered would have been closer to $R_V$ than $R_{\rm TA,~Max}$, and  such non-perturbative effect would have been reflected upon the motion of the members of a structure, residing just on the surface of it. There is no observational indication of such  peculiarilty of motion or `large' departure from geodesics, as far as the length scales of structures are concerned.

The extension of the turn around radius calculation to the next order of our perturbation scheme, by taking into account backreaction and $\mathcal O(\alpha^2)$ effects, is also important in order to confirm the stability of our conclusions. We hope to perform this analysis in the not too distant future.


\acknowledgments

This Research is implemented under the ``ARISTEIA II" Action of the Operational Program ``Education and Lifelong Learning" and is co-funded by the European Social Fund (ESF) and Greek National Resources. SB acknowledges this fund for partial support. TNT was partially supported by the European Union Seventh Framework Program (FP7-REGPOT-2012-2013-1) under grant agreement No. 316165.  SB is currently supported by his institution IUCAA, Pune, India. He thanks V.~Sahni for useful discussions. KFD acknowledges INFN Sez. di Napoli (Iniziativa Specifica TEONGRAV) for financial support. The authors would like to thank the anonymous referee for careful critical reading of the manuscript and for useful comments.



\begin{thebibliography}{99}

\bibitem{Martin:2012bt} 
  J.~Martin,
  Comptes Rendus Physique{\bf 13}, 566 (2012)
  [arXiv:1205.3365 [astro-ph.CO]].

\bibitem{polyakov}  A.~M.~Polyakov, arXiv:1209.4135[hep-th] and references therein.

\bibitem{ait} I.~Antoniadis, J.~Iliopoulos and T.~N.~Tomaras, Phys.\ Rev.\ Lett.{\bf 56},  1319 (1986).

\bibitem{woodard} R.~P.~Woodard, Int.\ J. \ Mod. \ Phys. D{\bf 23},  09, 1430020 (2014).

\bibitem{review}
T.~Clifton, P.~G.~Ferreira, A.~Padilla and C.~Skordis,
Phys.~Rep.~{\bf 513}, 1 (2012).

\bibitem{Sahni:2014ooa} 
  V.~Sahni, A.~Shafieloo and A.~A.~Starobinsky,
  Astrophys.\ J.\  {\bf 793}, no. 2, L40 (2014)
  [arXiv:1406.2209 [astro-ph.CO]].

\bibitem{Vikman}
C.~Deffayet, G.~Esposito-Farese, A.~Vikman,
 Phys.~Rev.~D{\bf79} 084003 (2009) [arXiv:0901.1314 [hep-th]].


	
\bibitem{deffayet} C.~Deffayet, G.~Esposito-Farese, D.~A.~Steer, Phys.~Rev.~D{\bf92} 084013 (2015) [arXiv:1506.01974[gr-qc]].



\bibitem{Martin-Moruno:2015kaa} 
  P.~Martin-Moruno and N.~J.~Nunes,
  JCAP{\bf 1509}, no. 09, 056 (2015)
  [arXiv:1506.02497 [gr-qc]].


\bibitem{Nicolis1}
A.~Nicolis, R.~Rattazzi, and E.~Trincherini, Phys. Rev. D{\bf 79}, 064036 (2009).

\bibitem{Dvali}
G.~Dvali,  G.~Gabadadze, and M.~Porrati, Phys. Lett. B{\bf 485}
208 (2000).

\bibitem{Luty}
M.~A.~Luty,  M.~Porrati,  and  R.~Rattazzi,  
JHEP{\bf 2003},  029 (2003).


\bibitem{Nicolis}
A.~Nicolis and R.~Rattazzi, 
JHEP{\bf 2004}, 059 (2004).

\bibitem{Rham}
C.~de~Rham,  Comptes ~Rendus ~Physique{\bf 13},  666  (2012).



\bibitem{Padilla:2010tj} 
C.~Deffayet, G.~Esposito-Farese and A.~Vikman,
 Phys.~Rev.~D{\bf79}, 084003 (2009)
[arXiv:0901.1314 [hep-th]]

\bibitem{babichev} E.~Babichev and C.~Deffayet,
 Class.\ Quant.\ Grav.~{\bf 30}, 184001 (2013).
 


\bibitem{Barreira:2013eea} 
  A.~Barreira, B.~Li, W.~A.~Hellwing, C.~M.~Baugh and S.~Pascoli,
  JCAP{\bf 1310}, 027 (2013)
  [arXiv:1306.3219 [astro-ph.CO]].


\bibitem{Tessore:2015sma} 
  N.~Tessore, H.~A.~Winther, R.~B.~Metcalf, P.~G.~Ferreira and C.~Giocoli,
  JCAP{\bf 1510}, no. 10, 036 (2015)
  [arXiv:1508.04011 [astro-ph.CO]].



\bibitem{Ali:2012cv} 
  A.~Ali, R.~Gannouji, M.~W.~Hossain and M.~Sami,
  Phys.\ Lett.\ B{\bf 718}, 5 (2013)
  [arXiv:1207.3959 [gr-qc]].



\bibitem{Hossain:2012qm} 
  M.~W.~Hossain and A.~A.~Sen,
  Phys.\ Lett.\ B{\bf 713}, 140 (2012)
  [arXiv:1201.6192 [astro-ph.CO]].
  

\bibitem{Bhattacharya:2015wlz} 
  S.~Bhattacharya, P.~Mukherjee, A.~S.~Roy and A.~Saha,
  arXiv:1512.03902 [gr-qc].


\bibitem{Hui:2012qt} 
  L.~Hui and A.~Nicolis,
  Phys.\ Rev.\ Lett.~{\bf 110}, no. 24, 241104 (2013).

\bibitem{Chris} E.~Babichev, C.~Charmousis and M.~Hassaine,
  JCAP{\bf 1505}, 031 (2015).

\bibitem{Iorio:2012pv}
  L.~Iorio,
  JCAP{\bf 1207}, 001 (2012)
  [arXiv:1204.0745 [gr-qc]].

\bibitem{Andrews:2013qva} 
  M.~Andrews, Y.~Z.~Chu and M.~Trodden,
  Phys.\ Rev.\ D{\bf 88}, 084028 (2013)
  [arXiv:1305.2194 [astro-ph.CO]].

\bibitem{Padilla:2010de} 
  A.~Padilla, P.~M.~Saffin and S.~Y.~Zhou,
  JHEP{\bf 1012}, 031 (2010)
  [arXiv:1007.5424 [hep-th]].


\bibitem{Padilla:2010tj} 
  A.~Padilla, P.~M.~Saffin and S.~Y.~Zhou,
  JHEP{\bf 1101}, 099 (2011)
  [arXiv:1008.3312 [hep-th]].


\bibitem{Vikman2}
C.~Deffayet, O.~Pujolas, I.~Sawicki and A.~Vikman
JCAP{\bf1010}, 026 (2010) [arXiv:1008.0048 [hep-th]]


\bibitem{Babichev:2012re} 
  E.~Babichev and G.~Esposito-Farese,
  Phys.\ Rev.\ D{\bf 87}, 044032 (2013).
  [arXiv:1212.1394 [gr-qc]].


\bibitem{Pavlidou:2013zha}
  V.~Pavlidou and T.~N.~Tomaras,
  JCAP{\bf 1409}, 020 (2014).
  
\bibitem{Tanoglidis:2014lea} 
  D.~Tanoglidis, V.~Pavlidou and T.~N.~Tomaras,
  arXiv:1412.6671; accepted for publication in JCAP.

\bibitem{Pavlidou:2014aia} 
  V.~Pavlidou, N.~Tetradis and T.~N.~Tomaras,
  JCAP{\bf 1405}, 017 (2014).
   
\bibitem{Bhattacharya:2015iha} 
  S.~Bhattacharya, K.~F.~Dialektopoulos, A.~E.~Romano and T.~N.~Tomaras,
 Phys.\ Rev.\ Lett.{\bf 115}, no. 18, 181104 (2015)
  arXiv:1505.02375 [gr-qc].  
  
\bibitem{Tanoglidis2015} D.~Tanoglidis, V.~Pavlidou and T.~N.~Tomaras; to appear.
  
\bibitem{Faraoni:2015zqa} 
V.~Faraoni, Phys.\ Dark Univ.~{\bf 11}, 11 (2015),  arXiv:1508.00475 [gr-qc]; V.~Faraoni, M.~Lapierre-Leonard and A.~Prain, JCAP {\bf 1510}, no. 10, 013 (2015), arXiv:1508.01725 [gr-qc].

\bibitem{Kaloper:2010ec} 
  N.~Kaloper, M.~Kleban and D.~Martin,
  Phys.\ Rev.\ D{\bf 81}, 104044 (2010)
  [arXiv:1003.4777 [hep-th]].


\bibitem{Silva}
F.~P~Silva and K.~Koyama,
Phys.\ Rev.\ D{\bf 80}, 121301 (2009).





\bibitem{Kastor:1992nn} 
  D.~Kastor and J.~H.~Traschen,
  Phys.\ Rev.\ D{\bf 47}, 5370 (1993)
  [hep-th/9212035].

\bibitem{Ratra}
B.~Ratra and P.~J.~E.~Peebles,
Phys.~Rev.~D{\bf37}, 3406 (1988)

\bibitem{Kunimitsu:2015faa} 
  T.~Kunimitsu, T.~Suyama, Y.~Watanabe and J.~Yokoyama,
  JCAP {\bf 1508}, no. 08, 044 (2015)
  [arXiv:1504.06946 [astro-ph.CO]].







\end{thebibliography}
\end{document}